\documentclass[pdflatex,sn-mathphys-num]{sn-jnl}% Math and Physical Sciences Numbered Reference Style
\usepackage{graphicx}%
\usepackage{multirow}%
\usepackage{amsmath,amssymb,amsfonts}%
\usepackage{amsthm}%
\usepackage{mathrsfs}%
\usepackage[title]{appendix}%
\usepackage{xcolor}%
\usepackage{textcomp}%
\usepackage{manyfoot}%
\usepackage{booktabs}%
\usepackage{algorithm}%
\usepackage{algorithmicx}%
\usepackage{algpseudocode}%
\usepackage{listings}%
\begin{document}

\title{Three Dimensional Structure of Jovian Polar Cyclones Revealed by Juno Microwave Radiometer}


\author[1]{\fnm{Jiheng} \sur{Hu}}%\email{jihenghu@umich.edu}
\author[2]{\fnm{Yuan} \sur{Lian}}%\email{lian@aeolisresearch.com}
\author*[1]{\fnm{Cheng} \sur{Li}}\email{chengcli@umich.edu}
\author[1]{\fnm{Sushil K.} \sur{Atreya}}%\email{atreya@umich.edu}
\author[3]{\fnm{Leigh N.} \sur{Fletcher}}%\email{leigh.fletcher@leicester.ac.uk}
% \author[3]{\fnm{Tristan} \sur{Guillot}}\email{tristan.guillot@oca.eu}

% \author[4]{\fnm{Tristan} \sur{Guillot}}\email{tristan.guillot@oca.eu}
\author[4]{\fnm{Yohai} \sur{Kaspi}}%\email{yohai.kaspi@weizmann.ac.il}
\author[5]{\fnm{Liming} \sur{Li}}%\email{lli7@central.uh.edu}
% \author[6]{\fnm{Alessandro} \sur{Mura}}\email{alessandro.mura@inaf.it}
\author[6]{\fnm{Glenn S.} \sur{Orton}}%\email{glenn.s.orton@jpl.nasa.gov}
\author[6]{\fnm{Fabiano A.} \sur{Oyafuso}}%\email{fabiano@jpl.nasa.gov}
\author[7]{\fnm{John H.} \sur{Rogers}}%\email{jrogers11@btinternet.com} % 0000-0002-4239-5907
\author[6]{\fnm{Fachreddin} \sur{Tabataba-Vakili}}%\email{fachtab@gmail.com}
\author[8]{\fnm{J. Hunter} \sur{Waite}}%\email{jhwaite@ua.edu}
\author[9,10]{\fnm{Michael H.} \sur{Wong}}%\email{mikewong@ssl.berkeley.edu} %0000-0003-2804-5086
\author[11]{\fnm{Zhimeng} \sur{Zhang}} %\email{zhimeng@caltech.edu}
\author[6]{\fnm{Steven M.} \sur{Levin}}%\email{steven.m.levin@jpl.nasa.gov}
\author[12]{\fnm{Scott J.} \sur{Bolton}}%\email{scott.bolton@swri.org}

\affil*[1]{Department of Climate and Space Sciences and Engineering, University of Michigan, Ann Arbor, MI, USA}
\affil[2]{Aeolis Research, Chandler, AZ, USA}
\affil[3]{School of Physics and Astronomy, University of Leicester, Leicester LE1 7RH, UK}

% \affil[3]{Université Côte d’Azur, OCA, Lagrange CNRS, Nice, France}
\affil[4]{Department of Earth and Planetary Sciences, Weizmann Institute of Science, Rehovot, Israel}
\affil[5]{Department of Physics, University of Houston, Houston, TX, USA}

% \affil[6]{INAF-Istituto di Astrofisica e Planetologia Spaziali, Roma, Italy}
\affil[6]{Jet Propulsion Laboratory, California Institute of Technology, Pasadena, CA, USA}
\affil[7]{British Astronomical Association, London, UK}
\affil[8]{Department of Physics and Astronomy, The University of Alabama, Tuscaloosa, AL, USA.}

\affil[9]{Space Sciences Laboratory, University of California, Berkeley, CA, USA}
\affil[10]{Carl Sagan Center for Science, SETI Institute, Mountain View, CA, USA}

\affil[11]{Department of Geological and Planetary Sciences, California Institute of Technology, Pasadena, CA, USA}
\affil[12]{Southwest Research Institute, San Antonio, TX, USA}


\abstract{ Jupiter’s circumpolar cyclones (CPCs) represent one of the most extreme dynamical regimes of the Solar System, where rotation, stratification, and convection interact to organize a stable polygonal cyclone lattice at each pole. 
Yet the depth and vertical structure of these cyclones remain unresolved.
Here, using Juno Microwave Radiometer observations acquired between May 2023 and May 2024, we retrieve the 3-D thermal/compositional structure of Jupiter’s northern CPCs at depths beyond those detectable by Earth-based telescopes and spacecraft that have ever visited.
The eight CPCs, appearing with enhanced microwave brightness, show as modest ammonia depletion near the water lifting-condensation level ($\sim$5 bar) and/or, equivalently, warm-core structures analogous to terrestrial tropical cyclones. Five of them show substantially stronger and deeper anomalies than the remaining three, with approximately twice the inferred ammonia depletion and localized heating.
In contrast, the central polar cyclone exhibits little microwave brightness contrast relative to its surroundings, despite its substantially cloudier and visually distinct appearance in visible and infrared observations. Previous dynamical models have treated the CPCs and the central polar cyclone as dynamically similar structures arising from the self-organization of turbulent flows. The new observations, however, suggest that the CPCs and the central polar cyclone may have distinct origins and operate under different dynamical regimes.}




\maketitle

NASA’s Juno mission revealed an intriguing polygonal arrangement of circumpolar cyclones (CPCs) surrounding a central polar cyclone at each of Jupiter’s poles \citep{adrianiClustersCyclonesEncircling2018, gavriel_yohai2021}. Markedly distinct from the transient polar vortices on terrestrial planets \citep{Mitchell2021-Review-PlanetaryPolarVortices}, Saturn’s hexagonal polar vortex  \citep{fletcherHexagonSaturnsNorthern2018, fletcherTemperatureCompositionSaturns2008}, and the alternating zonal-band structure at Jupiter’s lower latitudes \citep{smithJupiterSystemEyes1979, porcoCassiniImagingJupiters2003, ortonMultiplewavelengthSensingJupiter2017, fletcherJupitersTemperateBelt2021}, the CPCs raise fundamental questions about their formation and long-term stability. Addressing these questions may provide new insights into the coupling between Jupiter’s deep and shallow atmospheres, the mechanisms and distribution of energy injection, and the abundance of water at the pole as it regulates atmospheric stability\citep{oneillPolarVortexFormation2015, brueshaber2019, siegelmanMoistConvectionDrives2022, ShiJiaru2026,ingersollMoistConvectionEnergy2000, boltonMicrowaveObservationsReveal2021, Hu2026-hb}.

Several competing theories have been proposed to explain the formation and stability of the circumpolar cyclones (CPCs). Most treat the vortices within a single-layer framework, representing either the barotropic or first baroclinic mode of atmospheric motion, and attribute their formation to the inverse energy cascade associated with two-dimensional turbulence \citep{oneillPolarVortexFormation2015, brueshaber2019, liModelingStabilityPolygonal2020, Chen2025-hf, Gavriel2025-il, ShiJiaru2026}. Some of these reduced-order models have successfully reproduced the observed morphology, long-term stability, and drifting motions of the CPCs. However, direct comparisons with observations remain challenging because the primary observables are cloud morphology, infrared emission, and the spatial distributions of volatile species such as ammonia, rather than the dynamical variables predicted by a reduced order model. To date, the only three-dimensional model that reproduces the polygonal arrangement of Jupiter's circumpolar cyclones is that of Cai et al. \cite{caiDeepCloselyPacked2021}, which uses a fully compressible formulation with an enhanced internal heat flux and reduced gravity. Converting their nondimensional temperature perturbations into dimensional estimates suggests that the polar vortices are approximately 20–35 K colder than their surroundings and exhibit a predominantly barotropic, vertically coherent structure extending throughout the modeled convective atmosphere.

Microwave sounding provides a unique survey of the atmospheric thermal/compositional structures beneath the visible cloud deck.
Previous microwave observations from the Juno Microwave Radiometer (MWR), Very Large Array (VLA), and Atacama Large Millimeter/Submillimeter Array (ALMA) have shown that low- and mid-latitude features, such as the Great Red Spot (GRS), episodic storms, major belt disturbances, and folded filamentary regions, are associated with substantial variations in ammonia and water abundances, as well as thermal contrasts that extend far below the water condensation level \citep{depaterPeeringJupitersClouds2016, liDistributionAmmoniaJupiter2017, de_pater_first_2019, fletcherJupitersEquatorialPlumes2020, Guillot2020-br, Duer2021EvidenceMultipleFerrellike, boltonMicrowaveObservationsReveal2021, moeckelTempestsTroposphereMapping2025, Fletcher2026-FFR}. At the poles, however, polar cyclones were not identified prior to Juno, nor were they observed at microwave wavelengths by flyby missions or ground-based telescopes.

\begin{figure*}[ht!]
    \centering
    \includegraphics[width=0.85\linewidth]{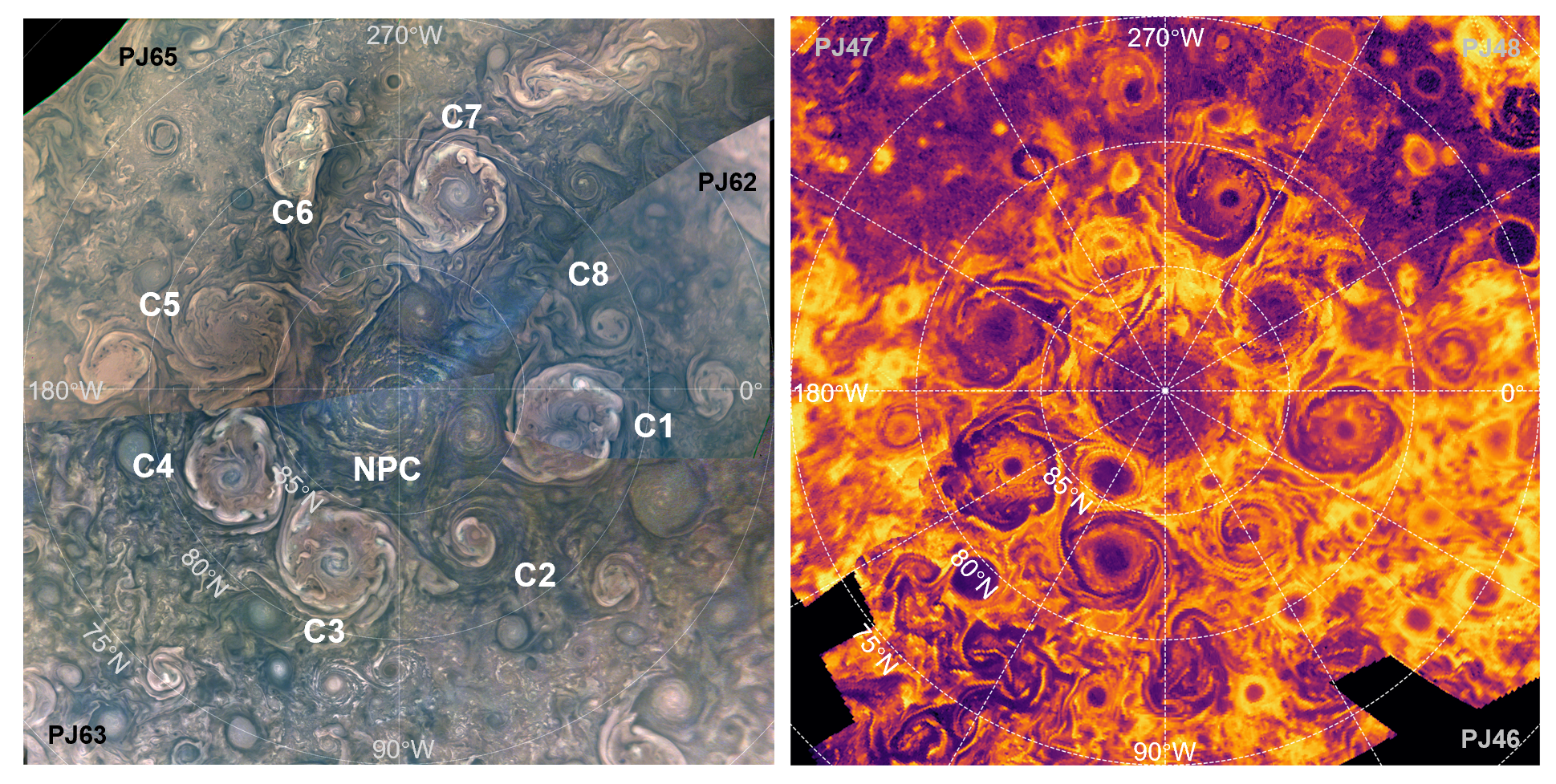}
    \caption{\textbf{Jupiter's north pole as it appears at visible and infrared wavelengths.}
    \textbf{Left panel}, JunoCam false color image, fused from observations during PJ62, PJ63, and PJ65 (Credit: NASA/JPL/SwRI/MSSS/Gerald Eichst\"adt/John Rogers). The north polar cyclone (NPC) and eight circumpolar cyclones (CPCs; ``C1''--``C8'') are labeled.
    \textbf{Right panel}, composite image of JIRAM’s 5-$\mu$m M-filter radiance, observed during PJ46, PJ47, and PJ48 (processed as described in \cite{Fletcher2026-FFR}). Both images are shown in polar orthographic projection (System III west longitude).
    }
    \label{fig:junocam_jiram}
\end{figure*}

The Juno MWR, designed with six centimeter-wave frequencies spanning 600 MHz (50 cm) to 22 GHz (1.36 cm), is able to probe both the vapor composition and the thermal structures beneath Jupiter’s cloud tops, reaching depths of up to $\sim$300 km \citep{janssenMWRMicrowaveRadiometer2017, boltonMicrowaveObservationsReveal2021}. 
With the advent of the 51st perijove pass (PJ51, 16 May 2023), Juno’s closest approach migrated poleward of 43$^\circ$N, while the spacecraft was descending to 4000 km above the visible cloud tops. This facilitates an unprecedentedly favorable observing geometry to retrieve the thermodynamic and compositional architectures of individual circumpolar cyclones and their surrounding environment at the Jovian north pole (see Fig.~\ref{fig:junocam_jiram}).

\section*{Microwave imagery of Jupiter's north pole}

\begin{figure*}[ht!]
    \centering
    \includegraphics[width=0.95\linewidth]{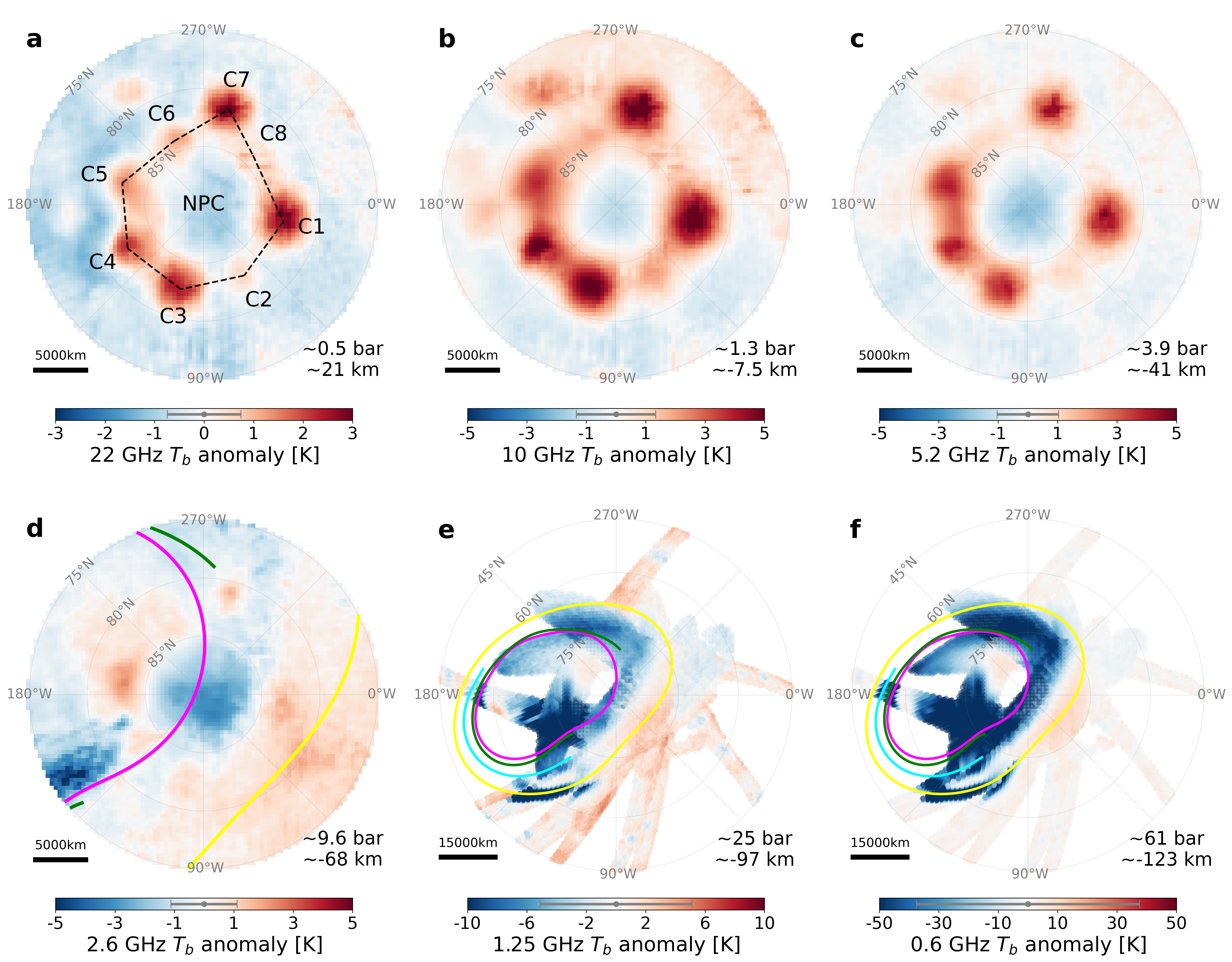}
    \caption{
    \textbf{Six-channel Juno/MWR brightness temperature ($T_\mathrm{b}$) anomaly maps over Jupiter’s north pole.}
    Panels \textbf{a--d,} Orthographic maps of $T_\mathrm{b}$ anomaly at 22, 10, 5.2, and 2.6~GHz relative to the polar-mean spectra (Table~\ref{tab:polarmean_spectra}), averaged over perijoves PJ51--PJ61 and gridded into Cartesian coordinates with a pixel resolution of $\sim$400~km. The dashed octagon in Panel \textbf{a} connects the centers of the eight CPCs.
    Panels \textbf{e--f,} Original 1.25 and 0.6~GHz $T_\mathrm{b}$ anomalies shown by perijove, extending to 45$^\circ$N to reveal the complete auroral ovals. In Panels \textbf{d--f}, we show the main auroral oval (magenta) and the footprints of Io (yellow), Europa (cyan), and Ganymede (light green). At both 0.6 and 1.25~GHz, a pronounced suppression of emission is observed, caused by cold emission from auroral electrons in the ionosphere and stratosphere \citep{bhattacharyaJupitersAuroralIonosphere2025}. The sampling pressure level and the corresponding altitude above the 1-bar pressure level are labeled for each frequency. Gray error bars superimposed on the color bars represent the range of polar area-weighted standard deviation ($\sigma$). They mostly represent atmospheric variations instead of measurement uncertainties.
    }
    \label{fig:anomaly_map}
\end{figure*}

The MWR antenna temperature measurements were collected over Jupiter’s north pole from PJ51 (May 2023) to PJ61 (May 2024), during which the circumpolar cyclones could be resolved by the MWR with high resolution. The raw data were subsequently converted to nadir brightness temperatures ($T_\mathrm{b}$) at each observation point through a series of data curation and spectral deconvolution procedures that consider both the instrumental viewing geometry and the angular dependence of the atmospheric emission (see Methods). The eight CPCs are generally stable in position and morphology \citep{tabataba-vakiliLongtermTrackingCircumpolar2020, rogers2021behaviour, muraOscillationsStabilityJupiter2021}. Conversely, weather features outside the octagon change rapidly from one perijove to the next. During a year, the CPCs did not move by more than 500 km in both meridional and zonal directions \citep{gavrielOscillatoryMotionJupiters2022, Gavriel&Kaspi2023TheWestwardDrift}. Therefore, they can be reliably represented by combining the MWR scans from different orbits. The spatial distribution of $T_\mathrm{b}$ anomalies relative to the polar-mean spectra (Table~\ref{tab:polarmean_spectra}) is presented in Fig.~\ref{fig:anomaly_map} for all six MWR frequencies, with a pixel resolution of approximately 400 km.

\begin{table}[h]
\centering
\caption{\textbf{Polar-mean nadir brightness temperature and limb darkening of Jovian north pole (75$^\circ$N$\sim$90$^\circ$N) during PJ51 to PJ61.}}
\label{tab:polarmean_spectra}%
\begin{tabular*}{\textwidth}{@{\extracolsep\fill}ccccc}
\toprule
Frequency  & Nadir $T_b$ & $R(45^\circ)$ & Sampling Pressure & Sampling Depth \\
(GHz) & (K) & (\%) & (bar) & (km) \\
\midrule
0.6   & 908.5$\pm$37.5   & 13.2 &  60.6 & -123.0 \\
1.25   & 477.7$\pm$5.1   & 10.0 & 25.1 & -96.5  \\
2.6   & 338.8$\pm$1.1   & 6.9 & 9.6 & -67.8 \\
5.2   & 251.7$\pm$1.0   & 5.9 & 3.9 & -40.7 \\
10   & 192.5$\pm$1.2   & 4.5 & 1.3 & -7.5 \\
22   & 141.1$\pm$0.7   & 1.4 & 0.5 & 21.2 \\
\bottomrule
\end{tabular*}
\footnotetext{$R(45^\circ)$ is limb darkening at emission angle of 45$^\circ$ (see Methods). Sampling pressure and depth refer to the pressure level and the corresponding altitude above the 1-bar level where the atmospheric thermal temperature is identical to the observed brightness temperature.}
\end{table}

At first glance, Jupiter’s north pole appears relatively serene, punctuated by eight CPCs that manifest as strong brightness temperature anomalies with diameters of approximately 4,000 to 5,000 km. 
In contrast to visible and infrared wavelengths (Fig.~\ref{fig:junocam_jiram}), the intricate morphologies of these cyclones, such as spiral-like extensions and wandering anticyclonic cores \cite{brueshaber2025morphological} observed in several CPC members, are absent from the microwave images due to the coarser spatial resolution and sensitivity to different opacity sources. 
However, the six microwave frequencies sound at different depths within Jupiter’s dense atmosphere. As shown in Fig.~\ref{fig:anomaly_map}, $T_\mathrm{b}$ anomalies of CPCs reach up to 3 K ($\sim$ 4$\sigma$) above the ambient atmosphere at 22 GHz, corresponding to a shallow sounding depth of $\sim$ 0.5 bar, which is conventionally considered to coincide with a cloud of ammonia condensate particles. $T_\mathrm{b}$ anomaly increases to approximately 5 K ($\sim$ 4$\sigma$) at 10 and 5.2 GHz, implying a substantial opacity and/or thermal disturbance around the water condensation layer at $\sim$ 7 bar. The anomalies gradually diminish at 2.6 GHz, indicating a relatively weak contrast between the cyclones and the ambient atmosphere at this altitude ($\sim$-60 km).

Measurements at 1.25 and 0.6 GHz are substantially contaminated by auroral emissions associated with energetic electron precipitation into the ionosphere and high-stratosphere \citep{bhattacharyaJupitersAuroralIonosphere2025}, preventing robust isolation of intrinsic atmospheric anomalies. Therefore, at the depth sensed by these two frequencies, the cyclonic disturbances may either merge smoothly into a more homogeneous interior state or remain independent but obscured by the auroral contamination. 
This is in contrast to the sign inversion in $T_\mathrm{b}$ anomalies between 2.6 and 5.2 GHz observed in a 38$^\circ$N cyclone \citep{boltonMicrowaveObservationsReveal2021} and high-latitude folded filamentary regions (FFRs) \citep{Fletcher2026-FFR}, where a drastic inversion in the ammonia or thermal anomaly structure is expected at around 5--7 bar level.

% $T_\mathrm{b}$ anomaly increases to approximately 5 K ($\sim$ 4$\sigma$) at 10 and 5.2 GHz, implying a substantial disturbance above the water condensation layer at $\sim$ 7 bar. The anomalies gradually diminish at 2.6 GHz, pronouncing a positive $T_\mathrm{b}$ anomaly that persists down to $\sim$8 bar ($\sim$60 km below the 1-bar level), before becoming obscured by substantial auroral contamination at 0.6 and 1.25 GHz \citep{bhattacharyaJupitersAuroralIonosphere2025}, preventing robust isolation of intrinsic atmospheric anomalies.
% Intriguingly, this behavior contrasts with the sign inversion in brightness temperature anomalies between 2.6 and 5.2 GHz observed in mid-latitude cyclones \citep{boltonMicrowaveObservationsReveal2021} and high-latitude folded filamentary regions \citep{Fletcher2026-FFR}. In those systems, the inversion implies a transition from an upper layer characterized by volatile depletion (and/or higher temperatures) to a deeper layer with enhanced volatile abundance (and/or lower temperatures), typically occurring around 5–7 bar, as predicted by theory \citep{Marcus2013JupiterRedOvalBA, Wong2011VerticalStructureOvalBA}. This suggests that in CPCs, either a microwave-dark anomaly at depth is absent, or the transition occurs at greater depths than in their low-latitude counterparts but is not observed due to auroral emissions.

The eight CPCs can be generally categorized into two groups, with cyclones C1, C3, C4, C5, and C7 appearing brighter in microwave imagery, whilst cyclones C2, C6, and C8 are relatively fainter. This is consistent with the two morphological classes of the eight CPCs demonstrated by JunoCam images (see Fig.~\ref{fig:junocam_jiram}a): the first five cyclones (since PJ12) are well-developed cyclones with `filled' morphology, while the remaining three are visually more diverse or amorphous \citep{tabataba-vakiliLongtermTrackingCircumpolar2020,brueshaber2025morphological}. 
Notably, the central cyclone, termed the north polar cyclone (NPC) by \cite{adrianiClustersCyclonesEncircling2018}, appears as a comparatively colder feature in microwave spectra, despite its substantially larger extent in visible and infrared imagery than the surrounding CPCs. This marked contrast implies that the NPC is a cold-core vortex and/or exhibits enhanced opacity from pressures of a few bars to the ammonia cloud top.

\section*{Vertical structure of polar cyclones from Bayesian inversion}
\begin{figure*}[hb!]
    \centering
    \includegraphics[width=1.0\linewidth]{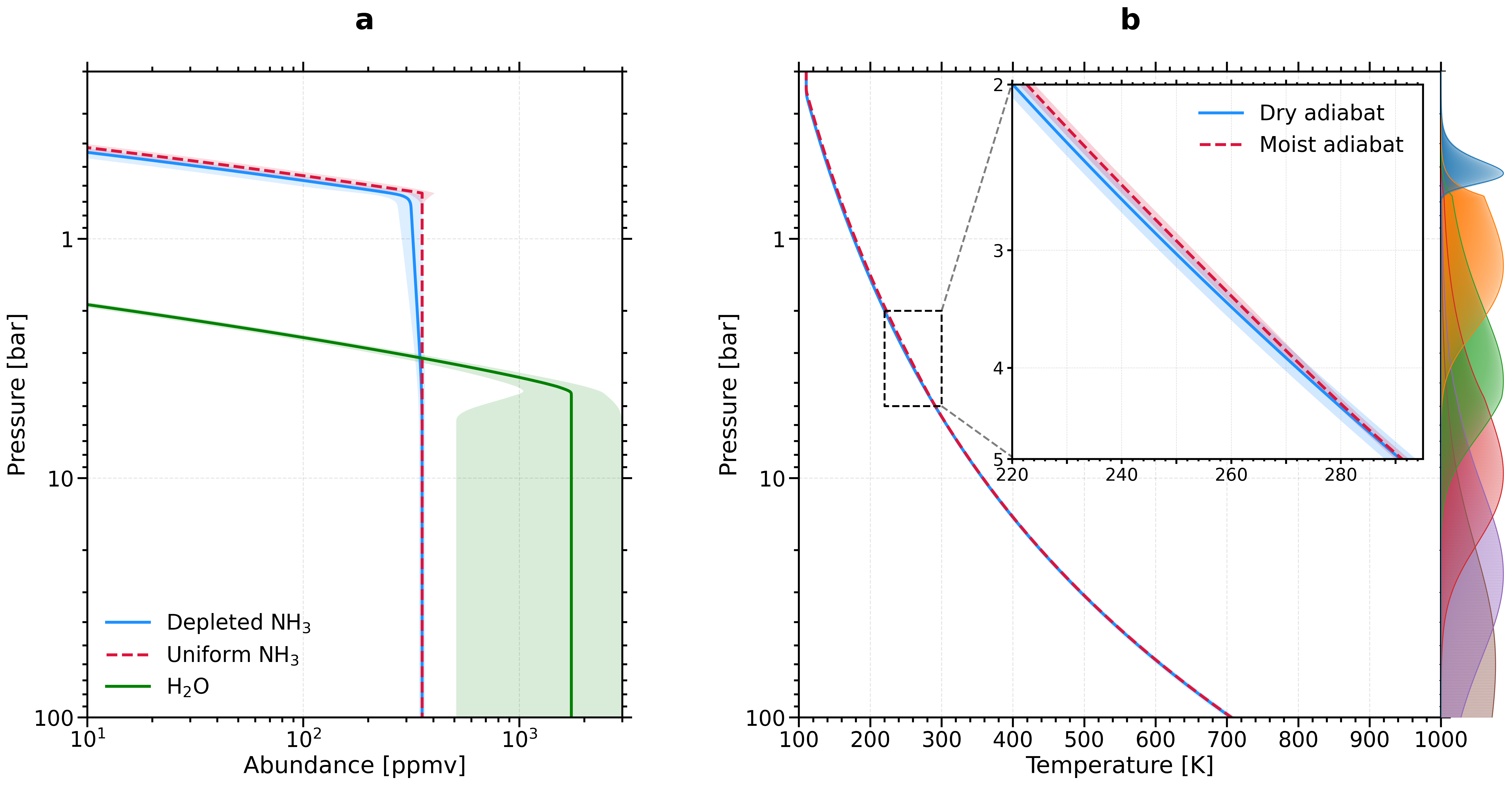}
    \caption{\textbf{Jovian north polar-mean atmospheric structures retrieved from Juno/MWR observations during PJ51 to PJ61.} Panel \textbf{a,} compositional abundances of water and ammonia vapor, for both a depleted and an uniform ammonia model, highlighting a different ammonia structure between 5 bar level to tropopause. Panel \textbf{b,} the thermal profiles under dry and moist adiabatic model, with the latter one being slightly warmer above the water cloud layer ($\sim$ 7 bar) due to localized latent heating. Colored histograms to the right of panel \textbf{b} are atmospheric weighting functions of the six MWR channels. To fit the Juno microwave observations, the dry adiabat was applied accompanying the depleted ammonia structure, while the moist adiabat was with the uniform ammonia profile. All atmospheric profiles are obtained from the medium values of the last 3,000 states of the chain advanced using the Markov chain Monte Carlo (MCMC) method. The shade denotes the envelop of 1$\sigma$ uncertainty. 
    }
    \label{fig:polar_background}
\end{figure*}

We retrieved atmospheric temperature and ammonia profiles for each pixel across the domain through a two-step inversion approach. First, we derived the polar-mean atmospheric profiles (Fig.~\ref{fig:polar_background}) using a Markov chain Monte Carlo (MCMC) approach \citep{foreman-mackeyEmceeMCMCHammer2013} constrained by the polar-mean $T_\mathrm{b}$ spectra (Table~\ref{tab:polarmean_spectra}) (see Methods). 
Using these polar-mean profiles as a prior state, we then applied a differential inversion to perform pixel-wise retrievals that reproduce the MWR observed brightness temperature anomalies (Fig.~\ref{fig:anomaly_map}). 

When retrieving the polar-mean atmospheric structure, we first test a baseline atmospheric state by assuming a dry-adiabatic temperature profile with an ammonia vapor profile uniformly mixed up to its cloud base. Under this assumption, the observed brightness temperature exceeds that predicted by the baseline model. We therefore consider two end-member scenarios that could produce the enhanced brightness temperature. In the first, the atmosphere follows a moist adiabat with homogeneous composition (a combination of red dashed lines in Figs.~\ref{fig:polar_background}a and \ref{fig:polar_background}b), yielding higher kinetic temperatures above the water-condensation level than in the dry-adiabatic case. In the second, the temperature profile remains dry adiabatic, but ammonia decreases with altitude, reducing absorption and increasing the brightness temperature (Fig.~\ref{fig:polar_background}, blue lines). We found that both polar-mean solutions match the observed $T_\mathbf{b}$ spectra well within the measurement uncertainty of a 2\% calibration error and a 0.5 K noise level. A detailed description of the MCMC inference for the polar-mean atmosphere is referred to \cite{Hu2026-hb} and the Methods. The resulting polar-mean atmospheric profiles serve as the benchmark state in the subsequent retrieval of pixel-wise cyclonic anomalies.

When performing pixel-wise differential retrieval, because atmospheric emissions at 0.6 and 1.25 GHz were not yet separated from other non-thermal emission sources, only the highest four MWR frequencies were used to constrain the cyclonic anomaly structures at altitudes no deeper than a prescribed pressure level $p_\mathrm{max}$ (see Methods). This limits the maximum value of $p_\mathrm{max}$ to approximately 10 bar, as constrained by the weighting function of the 2.5-GHz channel (Fig.~\ref{fig:polar_background}b, red histogram). 
At pressures lower than $p_\mathrm{max}$, we use three parameters to describe the atmosphere. One dimensionless parameter specifies the vertical gradient of ammonia vapor, $d \ln \mathrm{NH3}/ d\ln p$, up to the ammonia condensation level near 0.7 bar. A second parameter, proxied by the deep water abundance, represents the temperature anomaly produced by latent heat release above the water condensation level near 7 bar. A third parameter controls the relative humidity of ammonia above the 0.7 bar level (see Methods and Extended Data Fig. 1).

\begin{figure*}[ht!]
    \centering
    \includegraphics[width=\linewidth]{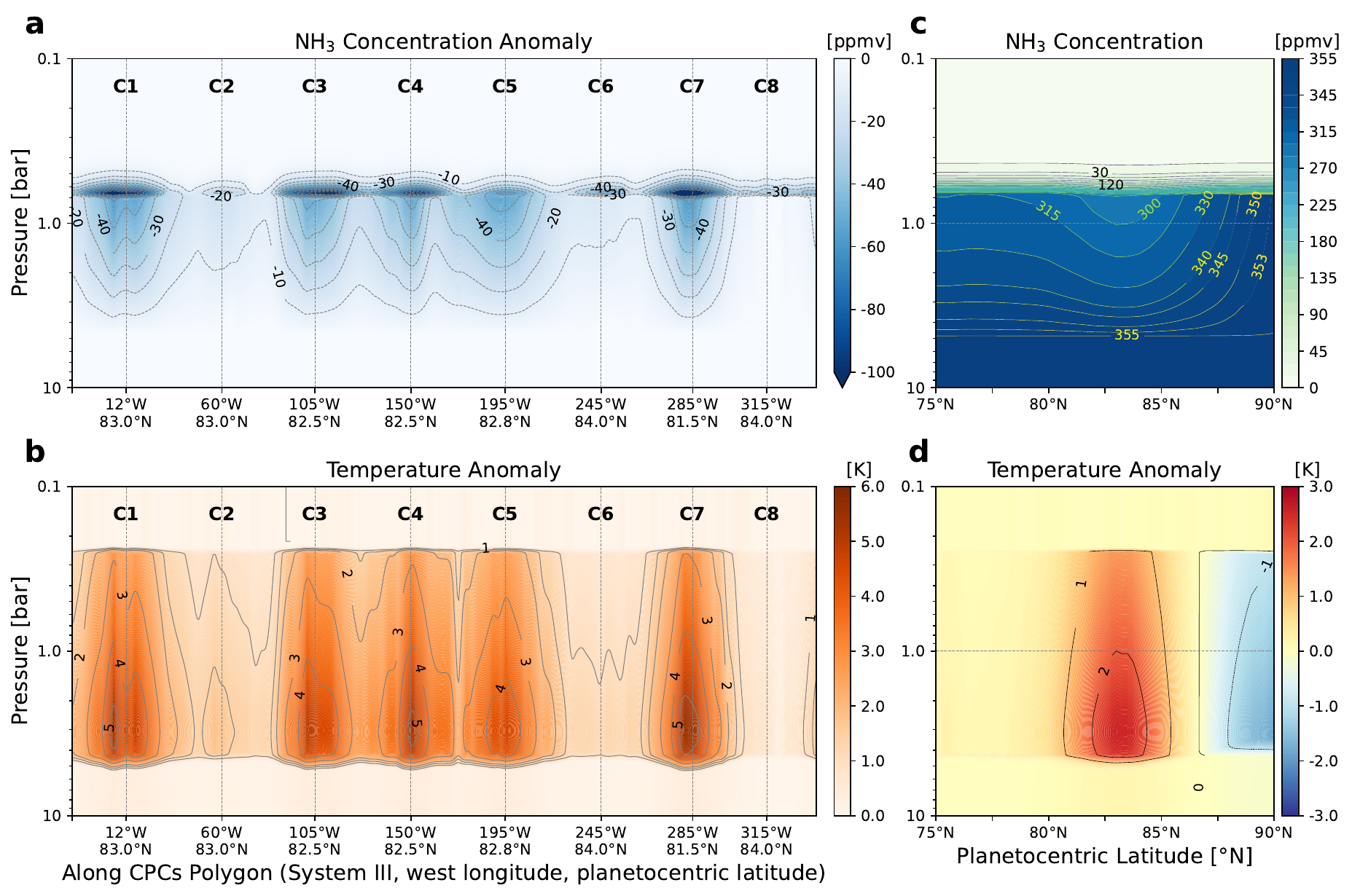}
    \caption{\textbf{Longitude/latitude-pressure cross sections of ammonia concentration and temperature}. Ammonia vapor concentration anomalies along the octagonal path (dashed polygon in Fig.~\ref{fig:anomaly_map}a) connecting the centers of the CPCs (Panel \textbf{a}) and zonal-mean ammonia vapor abundance from 75$^\circ$N to 90$^\circ$N (Panel \textbf{c}), are retrieved by fitting the measured $T_\mathrm{b}$ anomalies, perturbing only the ammonia structure while keeping the background temperature profile fixed.
    Significant depletion of ammonia is observed between levels of 0.5--5 bar at around 82.5$^\circ$N, as well as at the locations of each CPC. 
    Temperature anomaly profiles are shown along CPCs polygon (Panel \textbf{b}) and for the zonal mean (Panel \textbf{d}), retrieved by varying only the thermal structure while assuming ammonia to be in equilibrium. 
    Strong positive temperature anomalies dominate the 0.2–5 bar pressure range at the locations of the CPCs, whereas the NPC appears as an enclosure of colder area than the ambient air columns.
    }
    \label{fig:cross_section}
\end{figure*}

Fig.~\ref{fig:cross_section} shows the ammonia and temperature anomalies, relative to the polar-mean state, derived under these two scenarios along an octagonal path connecting the centers of the CPCs (Fig.~\ref{fig:anomaly_map}a, dashed polygon). In both cases, we find significant variability among individual CPCs; in particular, C1, C3, C4, C5, and C7 appear as stronger ammonia/thermal anomalies, whereas C2, C6, and C8 are relatively weaker, closely explaining the divergent brightness anomalies of these two categories of CPCs presented in Fig.~\ref{fig:anomaly_map}.

If interpreted as temperature anomalies, the five well-structured CPCs are warmer than the dry-adiabatic background by about 4--6 K, while the three amorphous CPCs are warmer by 1--2 K. If interpreted as compositional anomalies, the five strong CPCs are depleted in ammonia by approximately 40--100 ppmv ($\sim$10\%--25\% of the deep value), while the other three CPCs are weakly depleted by 20--30 ppmv. Both models indicate a similar vertical extent of the CPCs, with anomalies confined above the base of the water condensation level near 5 bar for the five stronger CPCs in both scenarios, whilst cyclones C2, C6, and C8 display weaker ammonia and thermal anomalies, and their vertical extent is shallower (down to the 2 bar level at most) in the ammonia anomaly. This depth should be regarded as only loosely constrained for two reasons: first, we did not use the 1.2 GHz data; second, the constructed profile is intentionally of minimal complexity, assuming a monotonic vertical gradient for either ammonia concentration or thermal contrast.

Figs.~\ref{fig:cross_section}c and ~\ref{fig:cross_section}d contrast the CPCs with the NPC by examining their zonal-mean structure. Despite their similar visual appearance in visible and infrared images, the CPCs differ markedly from the NPC in their microwave spectra and inferred atmospheric structure. Toward the pole, the zonal-mean ammonia abundance increases, or equivalently, the temperature decreases. In visible and infrared images, the NPC appears cloudy, as indicated by its higher visible reflectivity and lower infrared emission. However, in the microwave spectrum (Fig.~\ref{fig:anomaly_map}), the NPC is colder relative to the CPCs and is nearly indistinguishable from the surrounding polar background atmosphere. The inferred anomalies are therefore small, on the order of -1 K in temperature or +10 ppmv in ammonia abundance, which lies at the margin of the measurement noise level. Consequently, the NPC is interpreted to have a thermal and compositional structure similar to that of the background polar atmosphere.

\section*{Depth of the CPCs}

\begin{figure*}[ht!]
    \centering
    \includegraphics[width=0.95\linewidth]{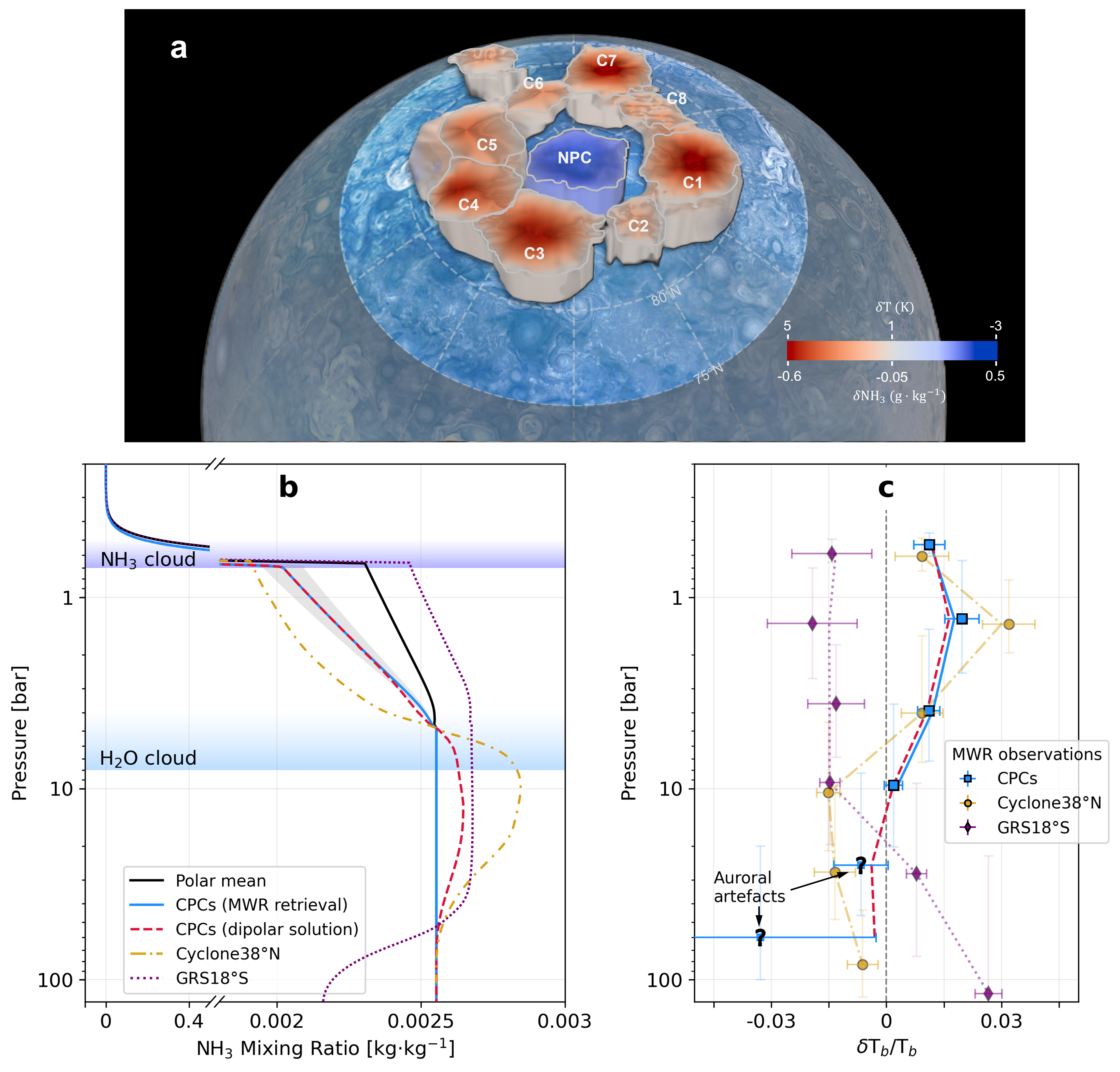}
    \caption{\textbf{The vertical structure of CPCs and the corresponding spectral $T_\mathrm{b}$ anomalies.}
    Panel \textbf{a} illustrates the three-dimensional CPC structure represented by simple atmospheric thermal and/or ammonia gradients. Panel \textbf{b} presents an alternative dipolar solution, with a thinner upper part overlying a denser lower part (red dashed line), as a trial-fit result. The blue line shows the gradient-depletion model retrieved in this study, rooted at 5 bar level. The profiles of a cyclone at 38$^\circ$N (dash-dotted line) and the Great Red Spot (GRS) at 18$^\circ$S (purple dotted line), retrieved from MWR observations during the PJ19 orbit \citep{boltonMicrowaveObservationsReveal2021}, are shown as ammonia anomalies superimposed on the polar-mean ammonia profile for comparison.
    Panel \textbf{c} presents the relative $T_\mathrm{b}$ anomalies observed in the CPCs, the 38$^\circ$N cyclone, and the GRS. Lines show the spectral fits to these observations using the corresponding profiles in panel \textbf{b}. \textbf{``?''} marks the two lowest MWR frequencies, which are contaminated by auroral emissions and therefore provide only loose constraints on the magnitude of the lower, colder branch of the proposed dipolar CPC structure.
    The CPC profiles and MWR spectra are averaged over the five strongest cyclones, C1, C3, C4, C5, and C7, to explore the deepest possible extent of CPCs.
    }
    \label{fig:CPC_depth}
\end{figure*}

The direct inversions described above represent the two simplest atmospheric structures over the 5 bar level that provide comparably good fits to the observed $T_\mathrm{b}$ anomalies of the highest four frequencies for each CPCs and NPC (Extended Data Fig. 2; see also the 3-D illustrations in Fig.~\ref{fig:CPC_depth}a). This degeneracy implies that the true atmospheric state may represent a combination of temperature and ammonia vapor anomalies in between these two end-member scenarios, and that microwave observations alone cannot uniquely distinguish between them. 
However, this approach does not exhaust all possible atmospheric configurations or constrain the true depth of the polar cyclones.

Inspired by the dipolar vertical structure of a cyclone at 38$^\circ$N revealed by the MWR spectra during PJ19 \citep{boltonMicrowaveObservationsReveal2021} (Fig.~\ref{fig:CPC_depth}b, golden dash-dotted line), we adapted a dipolar ammonia anomaly profile to characterize the vertical structure of CPCs (Fig.~\ref{fig:CPC_depth}b, red dashed line). Equivalent to the retrieved linear structure proposed above the 5-bar level, this dipolar structure provides an alternative solution that can reproduce the $T_\mathrm{b}$ anomalies at the four highest MWR frequencies, averaged over the five strongest CPCs (Fig.~\ref{fig:CPC_depth}c, red dashed line). Meanwhile, it allows for a deep reversed branch extending to depths well below the water condensation level.
Although this is a simple trial fit without constraints from the 0.6- and 1.25-GHz observations, the predicted $T_\mathrm{b}$ anomalies at these two channels fall near the lower bound of the $1\sigma$ uncertainty interval. Therefore, it is plausible that the stronger CPCs extend sufficiently deep to perturb the atmosphere at least to $\sim$60 bar level. The amplitude of the anomaly at depths below the water cloud layer, however, remains poorly constrained because reliable observations unaffected by auroral contamination are unavailable at 1.25 and 0.6 GHz.

Dynamically, cyclones are expected to have a warmer upper part over a high-density colder lower part to maintain hydrostatic and geostrophic balance, and vise versa for anticyclones \citep{Dowling1989-iv, Lemasquerier2020-my,Fletcher2026-FFR}. Accordingly, a deformation of the isentropic surfaces is expected, with a sign-inversion level located around levels of 5--8 bar, as observed in lower-latitude vortices, belts, and high-latitude FFRs \citep{boltonMicrowaveObservationsReveal2021, fletcherJupitersTemperateBelt2021, Fletcher2026-FFR}. 
However, in contrast to the PJ19 cyclone at 38$^\circ$N, whose $T_\mathrm{b}$ anomaly changes sign between 5.2 and 2.6 GHz (Fig.~\ref{fig:CPC_depth}c, golden-filled circles), the CPCs show no sign inversion even at 2.6 GHz (Fig.~\ref{fig:CPC_depth}c, blue-filled squares). 

The CPCs spectra may instead be more consistent with those of the well-known anticyclone Great Red Spot (GRS) at 18$^\circ$S (Fig.~\ref{fig:CPC_depth}c, purple-filled diamonds), whose $T_\mathrm{b}$ anomaly changes sign between 2.6 and 1.25 GHz, although the latter transitions from a colder upper branch to a warmer lower branch over the atmospheric depth probed by these two frequencies (Fig.~\ref{fig:CPC_depth}b, purple dotted line). This is also in line with the dynamical framework of \cite{Gavriel2025-il}, where a ``Mode 1'' structure predicts that the 1.25 GHz channel may detect a reversal in brightness temperature while remaining obscured in MWR signals by strong auroral and synchrotron contamination near the pole. The absence of a spectral inversion down to 2.6 GHz consolidates the argument that the CPCs have a deep, colder, high-density branch extending much deeper beneath Jupiter's weather layer.

\section*{Discussion}

Despite being compactly arranged and sharing some cyclonic characteristics, CPCs and NPC are suggested here to occupy distinct thermo-compositional regimes in the $\sim$0.6--7 bar layer: the CPCs are more strongly depleted in ammonia or locally warmer, whereas the horizontally larger NPC exhibits a weaker and cooler anomaly.

If the anomalies are primarily thermal, the contrast maps onto a familiar dichotomy in rotating atmospheres. Terrestrial polar vortices are large, radiatively maintained cold-core cyclones \citep{Charlton2007-da, holton2012introduction}, whereas cyclones energized by latent heating, such as tropical cyclones and Saturn's polar vortices, develop warm cores with compensating subsidence in the eye \citep{emanuel_air-sea_1986, dyudinaDynamicsSaturnsSouth2008, fletcherTemperatureCompositionSaturns2008}. In this framework, the NPC resembles the former, and the CPCs resemble the latter. %Thermal-wind balance then yields a testable prediction. A warm core between $\sim$0.6 and $\sim$5 bar requires the cyclonic winds of the CPCs to weaken with altitude across this layer, so cloud-tracked winds near the visible cloud deck \citep{Grassi2018, Ingersoll2022} would underestimate the circulation at the water-cloud level. The NPC, by contrast, should show weaker or reversed vertical shear. %
The warm CPCs cores suggested here do not contradict the predicted cold-core CPCs in the three-dimensional deep-convection simulations \citep{caiDeepCloselyPacked2021}, which are dry and represent the weather layer with Newtonian cooling. Instead, they indicate that the upper few bars of the CPCs are governed by moist processes absent from dry models.

The thermal and compositional interpretations need not be mutually exclusive. Sustained subsidence in a cyclone core would both import ammonia-poor air from above the ammonia lifting condensation level ($\sim$0.7 bar) and warm it adiabatically. The same coupling has been inferred from the warm, phosphine-depleted cores of Saturn's polar cyclones \citep{fletcherTemperatureCompositionSaturns2008, fletcher_thermal_2010}. Maintaining such subsidence against radiative cooling requires an energy source, most plausibly latent heat released in convective towers along the eyewall or in the surrounding flow. 
% JIRAM observations reveal vigorous moist convection among the polar cyclones that transfers energy upscale into the vortices \citep{Siegelman2022}. In addition, moist convective forcing combined with $\beta$-drift has been shown to accumulate cyclonic vorticity at giant-planet poles \citep{oneill2015}. 
Cloud microphysics provides further sinks for gaseous NH$_3$, through NH$_4$SH formation near 2--3 bar and dissolution into the water cloud near 5--7 bar \citep{Briggs1989-md}. Convective downdrafts and post-convective adjustment can then carry the depletion well below the cloud base \citep{liMoistConvectionHydrogen2015, Guillot2020-br}. 

% We therefore propose that the CPCs are actively sustained by moist convection, with their microwave signature recording time-integrated subsidence, whereas the NPC is a comparatively passive vortex closer to radiative control. One speculative explanation for this asymmetry is geometric: convectively generated cyclonic anomalies drifting poleward would encounter the ring of CPCs before reaching the pole \citep{gavriel2021}, leaving the NPC relatively starved of convective forcing.

% Two observations suggest that the microwave signature tracks forcing intensity rather than vortex size. First, the NPC is the largest vortex yet has the weakest anomaly. Second, t

The CPCs themselves differ remarkably: cyclones C2, C6, and C8 are weaker than the rest, despite occupying symmetric positions in the octagon. The geometry of the polygon is set by the balance between $\beta$-drift and the mutual repulsion of shielded cyclones \citep{liWaterAbundanceJupiters2020, gavriel_number_2021}, which constrains the potential-vorticity structure of the cyclones rather than their thermodynamic state. The weaker CPCs may therefore be at a different stage of a convective life cycle, consistent with the morphological variability and oscillations documented over the Juno mission \citep{muraOscillationsStabilityJupiter2021, gavriel_oscillatory_2022}.

% This decoupling between size and signal is itself useful. If vortex size and circulation strength are measured independently from imaging and cloud tracking, a dynamical model could predict the thermal anomaly of each cyclone; comparing that prediction with MWR would partly lift the temperature--ammonia degeneracy.

The vertical extent of the CPCs bears directly on their dynamics. Our retrieval primarily samples the upper portions of the vortices. However, the dipolar solution, with an ammonia-enhanced lower branch, suggests deeper roots. So does the absence, at frequencies above 2.5 GHz, of the delayed sign reversal expected from cyclonic balance \citep{boltonMicrowaveObservationsReveal2021, fletcherJupitersTemperateBelt2021,Fletcher2026-FFR}. Together, these points to a nearly isentropic mid-plane at several tens of bars, reminiscent of the GRS. Deep CPCs that survive in a convective environment would further require a stably stratified layer below the water condensation level, stabilized either by precipitation processes or by companion anticyclones \citep{Guillot2020-br, Guillot2020-wu}. Testing these ideas requires constraints not yet available: cyclone-specific limb darkening and signals in the 0.6- and 1.2-GHz channels from favorable polar geometries; JIRAM 5-$\mu$m spectra are sensitive to the few-bar level; and radio occultation constrains temperatures near 0.5 bar. Combined, these would separate thermal from compositional contributions and establish whether the NPC/CPCs dichotomy persists below the weather layer.

\section*{Methods}
\subsection*{Juno MWR observations}
NASA's Juno spacecraft completed eleven flyby observations over Jupiter's north pole from May 16, 2023, to May 12, 2024, with the perijove point of its highly elliptical orbit progressively marching poleward to 51$^\circ$N (PJ61) to facilitate close-in mapping of the polar region at favorable observing altitudes below 4000 km above the visible cloud top. During this period, the Juno microwave radiometer comprehensively sounded various weather systems using its six-frequency channels, spanning 600 MHz (Ch.1) to 22 GHz (Ch.6) \citep{janssenMWRMicrowaveRadiometer2017}, corresponding to the antenna beam width of approximately 20$^\circ$ for Ch.1 and Ch.2, and 10$^\circ$ for the four higher frequencies.
The raw measurements, antenna temperatures ($T_\mathrm{A}$), represent convolutions of the brightness temperatures ($T_\mathrm{b}$) emitted at different emission angles within the antenna pencil-beam field of view (FOV) \citep{oyafusoAngularDependenceSpatial2020, zhangResidualStudyTesting2020}. The convolution weights are determined by the antenna pattern and by the polar and azimuthal angles relative to the antenna bore-sight. $T_\mathrm{b}$ quantifies the integrated thermal emission from the Jovian atmosphere at the depth probed by the corresponding frequency, where the penetration depth depends on the atmospheric transmissivity and is, thus, a function of the vertical distribution of various opaque compositions.
In this paper, we collected $T_\mathrm{A}$ measurements within the north pole area (75$^\circ$N poleward), where the atmosphere is dominated by turbulent eddies and the polar vortices.

\subsection*{Spectra deconvolution}
Prior to any scientific interpretation, these data were subjected to a sequence of pre-processings documented by \cite{oyafusoAngularDependenceSpatial2020}, which, in summary, aim to screen for synchrotron emission, data corruption, lightning activity, and auroral influence. Only samples for which at least 85\% of the beam falls within an emission angle of $53^\circ$ were retained, which largely alleviates contamination from off-planet scenes and mitigates the beam-filling effect of varied emission angles.

An iterative deconvolution algorithm that considers both the emission angular dependence and the observational geometry \citep{zhangResidualStudyTesting2020} was applied to obtain the polar-mean spectra of $T_\mathrm{b}$. The method resembles the proposed latitudinal deconvolution method \citep{oyafusoAngularDependenceSpatial2020, liWaterAbundanceJupiters2020} that adopts a quadratic polynomial model to account for the angular dependence of $T_\mathrm{b}$ measured at the top of the atmosphere,
\begin{equation}\label{eq:tbmodel}
    {T_\mathrm{b}}_\lambda(\mu)=[a_\lambda+b_\lambda\cdot\mu+c_\lambda\cdot\mu^2]\cdot X(\mu)
\end{equation}
Here, $\lambda$ denotes the wavelength. $\mu=cos(\theta)$ is the cosine value of emission angle. Being functions of wavelength, $a$ represents the nadir ($\mu=1$) brightness temperature; $b$ and $c$ are parameters associated with limb darkening, which is defined as the brightness temperature darkening percentage at a given emission angle $\theta$ relative to the nadir view.

\begin{equation}\label{eq:Rtheta}
    R(\theta)=\frac{T_\mathrm{b}(0^\circ)-T_\mathrm{b}(\theta)}{T_\mathrm{b}(0^\circ)} \times 100\%
\end{equation}

In the context of radiative transfer in the Jovian atmosphere, both the nadir brightness temperature and the limb darkening are illuminating quantities of atmospheric opacity and temperature profiles.
Conventionally, $R(45^\circ)$ is used to characterize the limb darkening effect. $X(\mu)$ is an empirical shape scaling function introduced to better account for the additional drop-off in brightness temperature at large emission angles (i.e., $\mu<0.6$) \citep{oyafusoAngularDependenceSpatial2020}.

The deconvolution method is briefly summarized here. 
In the first step, the observed antenna temperatures, $T_\mathrm{A}$, from all perijoves (PJ51–PJ61) were collected over the polar region and screened to remove spurious contamination, including contributions from synchrotron radiation and auroral signals associated with high-density electron precipitations. The remaining $T_\mathrm{A}$ samples were then used to derive an initial estimate of the polar-mean brightness temperature model by fitting Eq.~\ref{eq:tbmodel} as a function of emission angle, yielding a first guess of the brightness temperature at each observation point, denoted as $T_\mathrm{b,model}$.
Next, for each emission angle bin, measurements deviating by more than one standard deviation from the median across all perijoves were excluded. The filtered samples were then refitted using Eq.~\ref{eq:tbmodel}, producing a robust estimate of $T_\mathrm{b,model}$ that represents the polar-mean brightness temperature model.

Within an iterative optimization framework, brightness temperatures, $T_\mathrm{b,model}$, were forward-modeled into antenna temperatures by convolving them with the MWR “pencil-beam” antenna gain pattern. Residuals between the modeled and observed $T_\mathrm{A}$ were evaluated at each observation point and then added to the $T_\mathrm{b}$ estimate to derive a new $T_\mathrm{b,model}$, yielding an updated estimate of the brightness temperature model that replaced the previous model derived from $T_\mathrm{A}$.
The forward convolution, residual evaluation, and update steps were repeated iteratively until convergence was achieved, defined as the point at which the polar-mean $T_\mathrm{b}$ (or equivalently, the residuals) no longer exhibited significant changes between successive iterations. This iterative retrieval procedure was applied independently to each MWR channel.

%In the first step, $T_\mathrm{A}$ observations from all perijoves from PJ51 to PJ61 were collected to drive a first guess of polar-mean $T_\mathrm{b}$ model coefficients by fitting the Eq.\ref{eq:tbmodel}, giving an initial guess of the mean polar brightness temperature for each observation point, $T_{b, fit}$; In an iterative optimal frame, $T_{b, fit}$s were convolved into a series of antenna temperature estimates, $T_{A, est}$, with the antenna gain pattern as the average kernel. Residual analysis was performed against the measured $T_\mathrm{A}$ values at each observation point to derive a set of updating coefficients, which were subsequently backpropagated to update $T_{b, fit}$. The iteration replicates the aforementioned procedure until the convergence of residuals is achieved. The iterative method was applied to each MWR channel separately. As reported, the residuals approached an instrumental noise of approximately 0.3 K for most observations in five to six iterations. 

Finally, the deconvolution method yields the polar-mean brightness temperature spectra (see Table~\ref{tab:polarmean_spectra} in Main-text). The spatial variability is characterized by the final residuals at each observation point, illustrating the thermal anomalies with respect to the polar-mean (see Fig.~\ref{fig:anomaly_map} in Main-text). %Data were organized into a two-dimensional orthogonal coordinate system, taking the north pole as the origin, which covers a field of 100$\times$100 pixels, with a uniform resolution of approximately 400 km. 

\subsection*{Bayesian inference of polar-mean atmospheric structure}
The polar-mean brightness temperatures at twenty channel-angles, that is, four emission angles (0$^\circ$, 15$^\circ$, 30$^\circ$, and 45$^\circ$) of five MWR channels from Ch.2 to Ch.6, were generated from Eq.~\ref{eq:tbmodel} and served as the observational constraints to retrieve the polar-mean atmospheric profiles. Ch.1 was not used due to the lack of accurate laboratory measurements of opacity and is severely affected by synchrotron radiation and auroral influences \citep{bhattacharya_highly_2023, aglyamov_alkali_2025, Zhang2026-AlkaliMineralDeepVariability}.

It was well demonstrated that the atmospheric structure over Jupiter’s north pole can be well characterized by a dry or moist adiabatic assumption \citep{liDistributionAmmoniaJupiter2017, liWaterAbundanceJupiters2020}, under which a collection of determinative parameters was designated to describe the regional mean vertical profiles of temperature and ammonia. The dry adiabatic model consists of five parameters: (1) deep-layer ($p>$ 8 bar) ammonia abundance, $x$NH$_3$, in parts per million by volume (ppmv), (2) deep layer potential temperature, $\theta_\mathrm{deep}$, (3) an upper limit on ammonia relative humidity, RH$_\mathrm{max}$, designed in particular to regulate the ammonia concentration at the NH$_3$-cloud layer, (4) an ammonia depletion gradient $\frac{d\text{lnNH}_3}{d\text{ln}p}$ (dimensionless), applied from (5) the $p_\mathrm{max}$ to 0.1 bar. As for the ideal moist adiabatic model, there are four parameters: (1) $x$NH$_3$, (2) $\theta_\mathrm{deep}$, (3) RH$_\mathrm{max}$, and (4) deep-layer water abundance in ppmv, $x$H$_2$O.

An equilibrium cloud condensation model \citep{liMoistAdiabatsMultiple2018, liSimulatingNonhydrostaticAtmospheres2019}, which accounts for the moist adiabatic processes in the atmospheres of giant planets, considers the condensation of multiple equilibrium gas species and temperature-dependent gas heat capacities, was used to construct the atmospheric structure according to the aforementioned thermodynamic scenarios and parameters. On top of that, radiative transfer simulations were performed to estimate the brightness temperature at the top of the visible cloud, considering the opacity of NH$_3$, H$_2$O, and the H$_2$/He collisional induced absorption (CIA). 

A Markov chain Monte Carlo (MCMC) sampler \citep{foreman-mackeyEmceeMCMCHammer2013} was used to explore the parameter space and assess the joint probability of these parameters by minimizing the departure of the simulated radiance from the observation at a given instrumental noise (0.5 K) and systematic error level (2\%). The prior input of the MCMC parameters was taken from the atmospheric retrievals in Jupiter's Equatorial Zone \citep{liWaterAbundanceJupiters2020}, namely, $x$NH$_3$=351$ _{-21}^{+22}$ ppmv, $x$H$_2$O=$2.5_{-1.6}^{+2.2}\times10^3$ ppmv, and $\theta_\mathrm{deep}$=169 K (referenced at 1 bar pressure level). Such a physically based variation scheme ensures that any alternative solution within the parameter space yields a convectively stable atmosphere, regardless of the adiabatic model applied. The resulting polar-mean profiles from the Bayesian retrievals are presented in Fig.~\ref{fig:polar_background}, for both dry and moist adiabat cases. 

Fig.~\ref{fig:polar_background}a presents the polar-mean ammonia vapor profile, assuming that the thermal structure follows the ideal dry adiabatic model. Constrained by the MWR spectra, deep ammonia is marginally determined as $x\text{NH}_3=354.8_{-11.0}^{+12.0}$ ppmv, corresponding to approximately $2.7\pm0.1$ times the solar O/H ratio. To fit the nadir $T_\mathrm{b}$ at 2.5 and 10 GHz, significant ammonia depletion around the levels of several bars is required, with a polar-mean gradient constrained to be $\frac{d\text{lnNH}_3}{d\text{ln}p}=-0.06\pm0.06$. This results in a total reduction of approximately -50 ppmv from the 5-bar pressure level to the ammonia cloud deck, leading to an abundance of approximately 300 ppmv at the 1-bar level. As another end member, employing an ideal moist adiabatic model, the polar-mean atmospheric temperature profile (Fig.~\ref{fig:polar_background}b, red dashed line) depicts localized heating from the 5 bar level to the tropopause, attributed to the latent heat release of water vapor, with a deep-layer abundance inferred to be $1.8_{-1.1}^{+1.5}\times10^3$ ppmv ($2.0_{-1.2}^{+1.7} \times$ Solar). RH$_\mathrm{max}$ for ammonia was inferred to be $0.71\pm0.15$ for both adiabatic cases. The deep potential temperature was inferred to be $\theta_\mathrm{deep}$=175.7$\pm$1 K.
We refer readers to \cite{Hu2026-hb} for detailed documentation of polar-mean atmospheric retrievals.

\subsection*{Pixel-wise differential inversion of cyclonic structures}\label{diffmethods}

In an attempt to determine the pixel-wise atmospheric structures, the polar-mean atmospheric profiles were fixed as benchmarks, and a differential inversion method was adopted for each pixel across the domain to retrieve their anomaly structures with respect to the benchmark profiles, aiming to reproduce the MWR observed brightness temperature anomalies relative to the polar-mean spectra (Fig.~\ref{fig:anomaly_map}). 
% The method can substantially eliminate the uncertainty induced by instrumental calibration errors, which are of comparable magnitude to the lateral thermal anomalies in relatively quiescent regions.

Inherited from the polar-mean benchmarks, the differential structures also have two end-member solutions based on two distinct thermodynamic assumptions.
% Because only the highest four MWR frequencies were used to retrieve cyclonic anomaly structures, we fixed the deep ammonia abundance ($x${NH}$_3$) and deep potential temperature ($\theta_\mathrm{deep}$) for both thermal end-members and only varied the ammonia gradient ($\frac{d\text{lnNH}_3}{d\text{ln}p}$, applied to dry adiabatic only), a temperature gradient above 7 bar due to latent heat associated with water condensation (proxied by the deep water abundance $x$H$_2$O), and an upper-limit on ammonia relative humidity (RH$_\mathrm{max}$). 
When tuning the profiles, the polar-mean deep ammonia abundance ($x\text{NH}_3$) and deep potential temperature ($\theta_\mathrm{deep}$) were applied directly to each pixel under both thermodynamical end-members, due to the fact that the frequencies used for the anomalous retrievals are larger than 2.6 GHz, and frequencies smaller than this value are not sensitive to pressures lower than 8 bar.
We only vary parameters that mediate shallow ammonia/thermal structures, namely, the ammonia gradient ($\frac{d\text{lnNH}_3}{d\text{ln}p}$, applied to dry adiabat only), a temperature gradient above the level of 7 bar due to latent heat associated with water condensation (proxied by the deep water abundance $x$H$_2$O, applied to moist adiabat only), and an upper-limit on ammonia relative humidity (RH$_\mathrm{max}$, for both cases). 
The error covariance matrices of these parameters were adopted from the corresponding posterior probability distribution inferred in the polar-mean atmosphere inversion. The spatial distributions of these parameters resulting from differential inversions are presented in Extended Data Fig. 1.

The differential inversion method, in which the optimal profile was derived to reproduce the observed anomalies relative to the polar-mean rather than the absolute value of brightness temperatures, was employed. This approach effectively mitigates the uncertainty associated with systematic observational errors and modeling errors while focusing on characterizing the structure of each atmospheric column, particularly within polar cyclones. For each air column, a cost function, $J(\mathbf{x})$, which quantifies the mismatch between simulated and observed anomalies, is minimized.

% \begin{align*}
% (x+y)^3&=(x+y)(x+y)^2\\
%        &=(x+y)(x^2+2xy+y^2) \numberthis \label{eqn:example1} \\
%        &=x^3+3x^2y+3xy^3+x^3.
% \end{align*}

\begin{align}
J(\mathbf{x}) = &\frac{1}{2} \left[ \mathbf{y}'_{\mathrm{sim}}(\mathbf{x}) - \mathbf{y}'_{\mathrm{obs}} \right]^{\mathrm{T}} \mathbf{S}_\epsilon^{-1} \left[ \mathbf{y}'_{\mathrm{sim}}(\mathbf{x}) - \mathbf{y}'_{\mathrm{obs}} \right] \\
&+\frac{1}{2} \left[ \mathbf{x} - \mathbf{x}_a \right]^{\mathrm{T}} \mathbf{S}_a^{-1} \left[ \mathbf{x} - \mathbf{x}_a \right],
\end{align}
Here, $\mathbf{x}$ is the state vector (e.g., temperature and gas concentration profiles); $\mathbf{x}_a$ is the prior state (polar-mean profile); $\mathbf{y}'_{\mathrm{obs}}$ and $\mathbf{y}'_{\mathrm{sim}}(\mathbf{x})$ are the observed and simulated anomalies relative to the polar-mean spectra, respectively. $\mathbf{S}_\epsilon$ and $\mathbf{S}_a$ are the observational and prior error covariance matrices. In the differential inversion process, systematic errors are effectively eliminated by the double-difference formulation, leaving only random noise to penalize the mismatch between simulated and observed $T_\mathrm{b}$ anomalies. Consequently, $\mathbf{S}_\epsilon$ becomes a diagonal matrix, where each diagonal element represents random noise with a value of 0.5 K, estimated from the deconvolution analysis \citep{zhangResidualStudyTesting2020}. We adopt an iterative limited-memory Broyden-Fletcher-Goldfarb-Shanno approach with box constraints (L-BFGS-B) \citep{zhu1997-L-BFGS-B} to minimize the cost function \( J(\mathbf{x}) \) and achieve the optimal state estimate \( \mathbf{x} \). This approach is computationally more efficient than an MCMC-based retrieval and is therefore more suitable for large-scale applications involving 10$^4$ pixels in this study.

\backmatter

\bmhead{Acknowledgements}
J.H. and C.L. are supported by the NASA Juno Program, under NASA Contract NNM06AA75C from the Marshall Space Flight Center, through subcontract 699056KC and Q99063JAR to the University of Michigan from the Southwest Research Institute.
Some of this research was carried out at the Jet Propulsion Laboratory, California Institute of Technology, under a contract with the National Aeronautics and Space Administration (80NM0018D0004). 
Y.L. was supported by the NASA Juno PSP under grant number 80NSSC22K0370.
L.N.F. is supported by the Science and Technology Facilities Council Consolidated Grant reference UKRI1205. M.H.W. is supported by the NASA New Frontiers Data Analysis Program (80NSSC25K0362).

\bmhead{Data and code availability}
Data and code will be made available upon publication.
\bmhead{Declarations}
The authors declare no conflict of interest.

\clearpage
\begin{appendices}
\section*{Extended Data}\label{secA1}

\renewcommand{\figurename}{}
\renewcommand{\thefigure}{Extended Data Fig.~\arabic{figure}}

\renewcommand{\tablename}{}
\renewcommand{\thetable}{Extended Data Table~\arabic{table}}

\begin{figure*}[htb]
    \centering

    \includegraphics[width=0.9\linewidth]{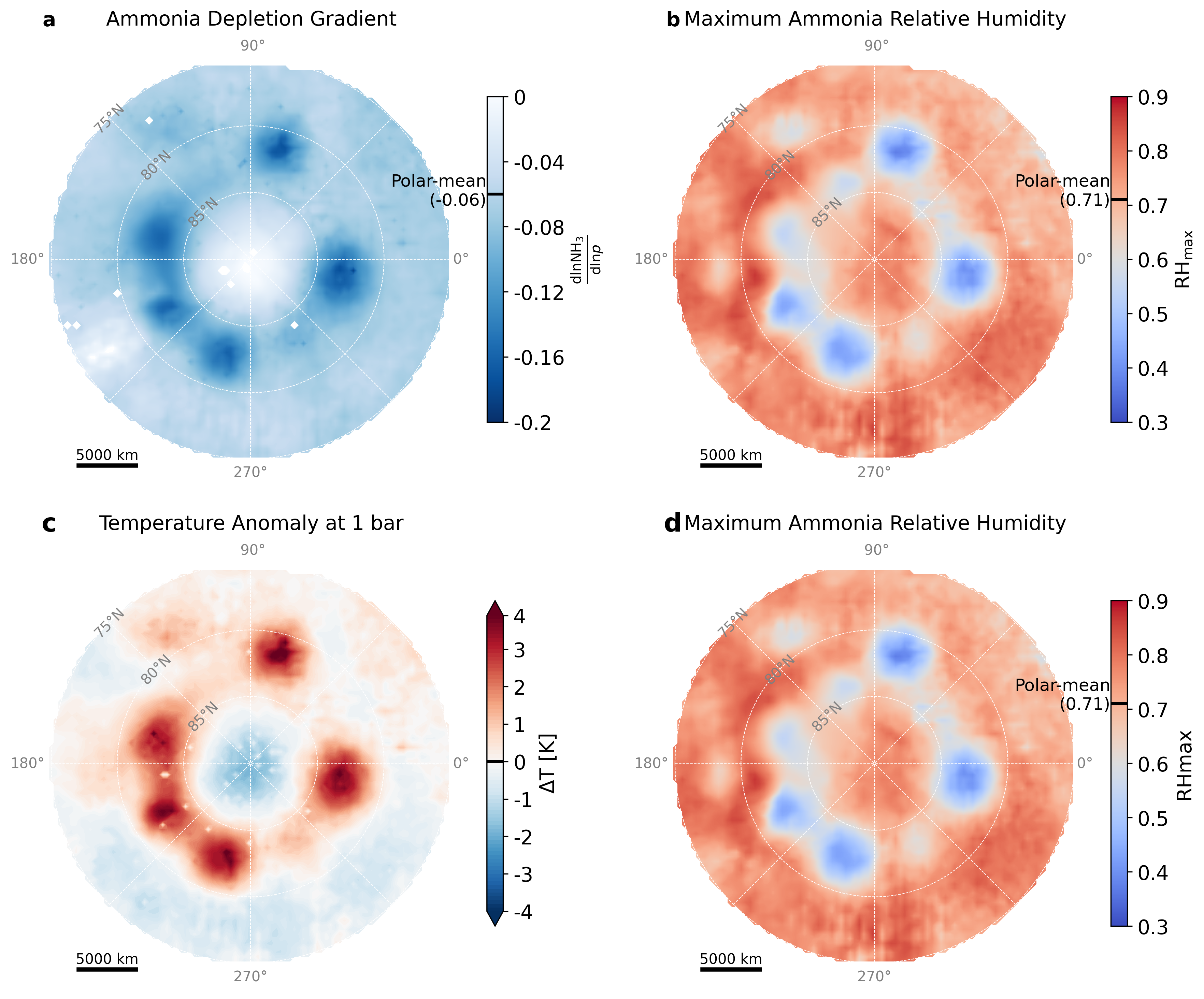}
    \caption{\textbf{The inferred parameters of two atmospheric structure end-members from differential inversion method.} The ammonia depleted gradient (Panel~\textbf{a}) and the maximum ammonia relative humidity (Panel~\textbf{b}) of the dry adiabatic solution. The temperature anomalies at 1 bar level (Panel~\textbf{c}) and the maximum ammonia relative humidity (Panel \textbf{d}) for the moist adiabatic solution. Blank pixels represent pixels where the differential inversion method failed to converge. Black bars in the color bars denote the polar-mean values used as base state for retrieval.}
    \label{fig:dry_params}
\end{figure*}
\clearpage

\begin{figure*}
    \centering
    \includegraphics[width=0.95\linewidth]{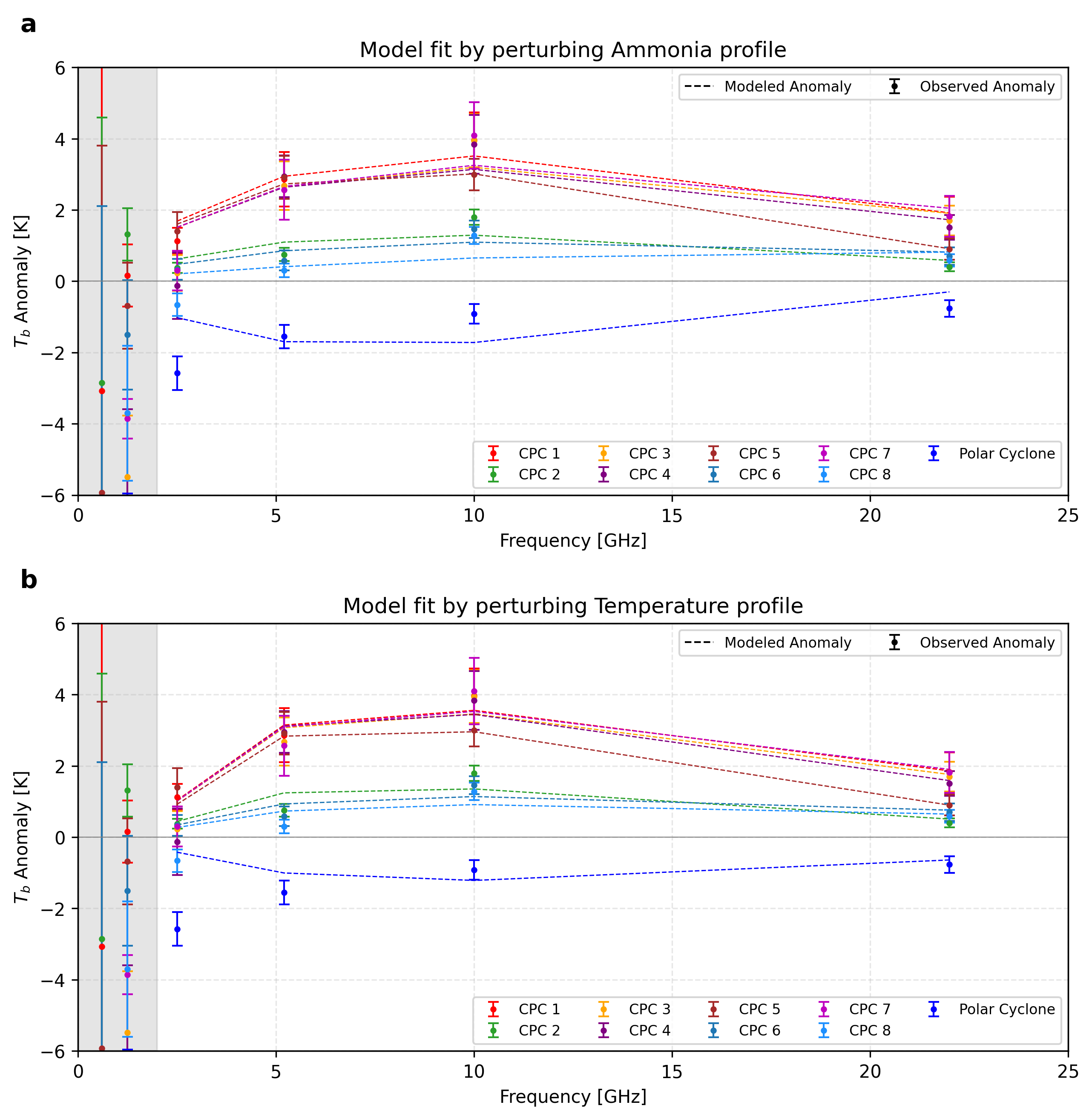}
    \caption{\textbf{The observed and modeled brightness temperature anomalies for each individual polar cyclone, relative to the polar background.}
    Panel~\textbf{a,} Modeled anomalies resulting from perturbed ammonia structures, assuming an undisturbed thermal structure.
    Panel~\textbf{b,} Modeled anomalies resulting from perturbed temperature structures, assuming an unperturbed composition field.
    For both cases, dots with 1$\sigma$ error bars represent the observed anomalies spatially averaged within each cyclone. Dashed lines indicate the corresponding modeled anomalies. Both models reproduce the observations well at the highest four frequency channels for both the north polar cyclone (NPC) and the circumpolar cyclones (CPCs). Observations at 0.6 and 1.25 GHz are significantly contaminated by auroral emissions (shaded region in both panels), therefore were excluded from fitting in the inversions.}
    \label{fig:model_fits}
\end{figure*}
\clearpage

\end{appendices}


% Bibliography
\bibliography{paperpile,references,sn-bibliography}

\end{document}